\documentclass[preprint,showpacs,preprintnumbers,nofootinbib,amsmath,amssymb,showkeys]{revtex4}

\usepackage{graphicx}
\usepackage{amsmath}

\begin{document}

\title{Theoretical analysis of telescopic oscillations in multi-walled carbon nanotubes}

\author{Vladimir Zavalniuk}
\email[]{VZavalnyuk@onu.edu.ua}%
\author{Sergey Marchenko}
\email[]{Sergey.Marchenko@onu.edu.ua}%
\affiliation {Department of Theoretical Physics, Odessa I.I.Mechnikov National University, \\ 2 Dvoryanska St., Odessa 65026, Ukraine}


\begin{abstract}
A simplified theory of the telescopic oscillations in multiwalled carbon nanotubes is developed. The explicit expressions for the telescopic force constants (longitudinal rigidity) and the frequencies of telescopic oscillations are derived. The contribution of small-amplitude telescopic oscillations to the nanotubes low temperature specific heat is estimated.
\end{abstract}

\maketitle

\section{Introduction}

Multi-walled carbon nanotubes (MWCNTs) are the first discovered nanoscopic quasi-1D nanostructures \cite{iijima}. Each MWCNT consists of some
nested single-walled nanotubes (shells) held mostly by van der Waals forces \cite{girifalco}.

The telescopic motion ability of inner shells \cite{zettl} and their unique mechanical properties \cite{treasy} permit to use
multi-walled nanotubes as main movable arms in coming nanomechanical devices. The variety of gadgets of this kind was already suggested such as
a possible mechanical gigahertz oscillator (linear bearing) \cite{zettl, zheng}, nanoswitch \cite{forro}, nanorelay and nanogear
\cite{strastava}, nanorail,  reciprocating nanoengine \cite{tu}. Therefore the analysis of mechanical characteristics of MWCNT is an important
objective of study. The present work is devoted to a simplified continuum version of this problem. The continuum model for telescopic
oscillations, in which each shell of MWCNT is considered as continuous infinitesimally thin cylinder is described in the next section. The
third section is devoted to the description of the small (thermal) and large-amplitude oscillations for DWCNT and MWCNT in the framework of
proposed model. Note that the similar continuum model was used recently for the investigation of the suction energy and large amplitude telescopic oscillations in DWCNT \cite{baowan1, baowan2}. The contribution of temperature-induced oscillations into the tubes heat capacity within Debye model is also discussed. In the last section the obtained results are compared with the available experimental data \cite{zettl}.

\section{Intertube interaction in MWCNT within continuum model}

The interaction energy of two shells of the multi-walled tube is modelled as the sum of pair interaction potentials of atoms from different shells.
In doing so we took for the potential energy of two atoms at the distance $l$ the Lennard-Jones potential
$$E_{LJ}(l)=-\frac{\gamma_{6}}{l^{6}}+\frac{\gamma_{12}}{l^{12}},$$
with attractive and repulsive constants $\gamma_{6}=2.43\times10^{-24} \,\, \textrm{J}{\cdot}\textrm{nm}^{6}$ and $\gamma_{12}=3.859\times10^{-27} \,\,
\textrm{J}{\cdot}\textrm{nm}^{12}$ borrowed from \cite{girifalco}. In accordance with this approximation the total intertube interaction energy takes the form
\begin{equation}\label{1.2}
 U=\sum_{i=1}^{N_{1}}\sum_{j=1}^{N_{2}}\left(-\frac{\gamma_{6}}{(\textbf{r}_{1,i}-\textbf{r}_{2,j})^6}
 +\frac{\gamma_{12}}{(\textbf{r}_{1,i}-\textbf{r}_{2,j})^{12}}\right),
\end{equation}
where $\textbf{r}_{1,i}$ and $\textbf{r}_{2,j}$ are radii vectores of the inner and outer tube's atoms respectively.

As in \cite{girifalco} we used instead of (\ref{1.2}) the continuum model, for which
\begin{multline}\label{1.3}
 U(\Delta z)=\sigma^2 r_{1} r_{2}\int\limits_{0}^{2\pi}
 d\theta_{1}\int\limits_{0}^{2\pi}d\theta_{2}
 \int\limits_{L_2-L_1+\Delta z}^{L_2+\Delta z}dz_{1}\int\limits_{0}^{L_2}dz_{2}\left(
 \frac{\gamma_{12}}{(r_{1}^{2}+r_{2}^{2}-2r_{1}r_{2}\cos(\theta_{1}{-}\theta_{2})+(z_{1}{-}z_{2})^2)^{6}}
 \right. \\
 \left.-\frac{\gamma_{6}}{(r_{1}^{2}+r_{2}^{2}-2r_{1}r_{2}\cos(\theta_{1}-\theta_{2})+(z_{1}-z_{2})^2)^3}\right),
\end{multline}
where $r_1$ and $r_2$ are inner and outer tubes radii, $L_1$ and $L_2$ are their lengths (from now on we assume that $L_1\leq L_2$) and $\Delta
z$ is the distance between tubes outer edges and $\sigma$ is the surface density of carbon atoms in graphene, which is almost independent on the
tube chirality,
$$\sigma=\frac{4}{3 \sqrt{3} \,\, b^{2}}=38.2 \, \textrm{nm}^{-2},$$
where $b=0.142 \, \textrm{nm}$ is the interatomic distance in graphene. Note that expression (\ref{1.3}) governs any one of coaxial DWCNT configurations, but for stable natural multiwalled nanotubes the interlayer distance $d$ ranges from 0.342 to 0.375 nm, and that it is a function of the curvature \cite{kiang}.

The integration over variables $z_{1}$ and $z_{2}$ can be easily carried out analytically, but obtained expressions are too cumbersome to be presented here.

It's clear, that the system energy is minimal when the inner tube is completely retracted into the outer tube. In terms of hypergeometric
functions the minimum interaction energy is given by expression
\begin{multline}\label{1.5}
U_{min}=\frac{3}{2} \pi^{3} \sigma^{2} r_{1} r_{2} \min(L_1,L_2)\times
\\
\times \left( \frac{21}{32} \, \gamma_{12} \frac{\,_{2}\textrm{F}_{1} \left( \frac{1}{2}, \frac{11}{2}, 1, \frac{4r_{1}r_{2}}{(r_{1}+r_{2})^{2}} \right)}{(r_{1}+r_{2})^{11}} -\gamma_{6} \frac{\,_{2}\textrm{F}_{1} \left( \frac{1}{2}, \frac{5}{2}, 1, \frac{4r_{1}r_{2}}{(r_{1}+r_{2})^{2}} \right)}{(r_{1}+r_{2})^{5}} \right),
\end{multline}
where
$$
\,_{2}\textrm{F}_{1} \left( \frac{1}{2}, J, 1, \frac{4r_{1}r_{2}}{(r_{1}+r_{2})^{2}} \right) =
\frac{(r_{1}+r_{2})^{2J}}{2\pi} \int_{-\pi}^{\pi}\frac{d \theta}{(r_{1}^{2}+r_{2}^{2}-2r_{1}r_{2}cos\theta)^{J}}.
$$

\begin{figure}[!hbp]
\includegraphics[scale=0.13]{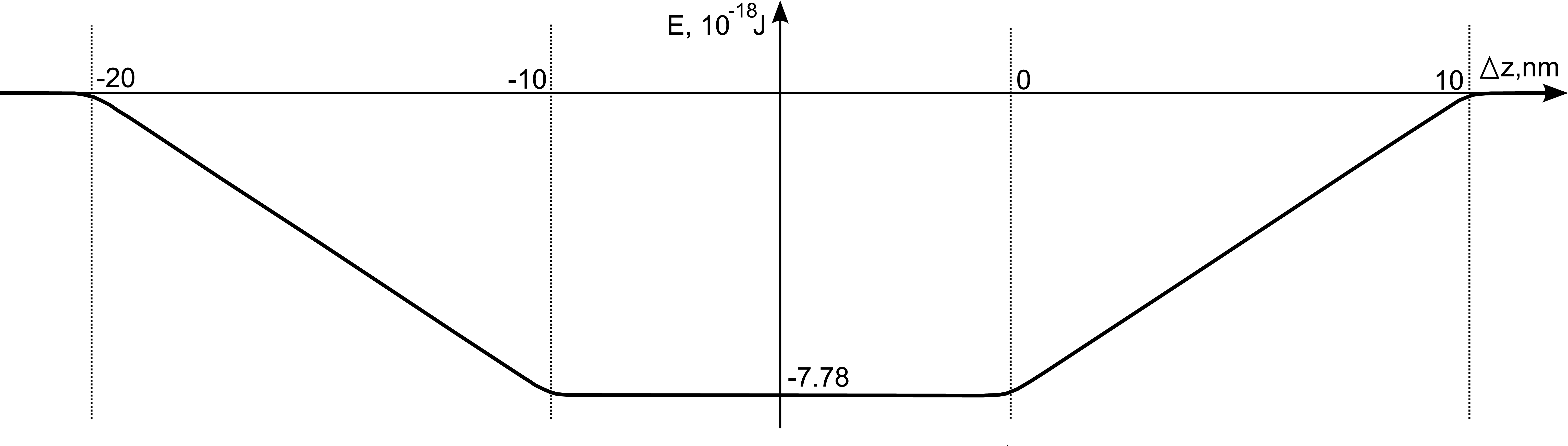}
\caption{The intertube interaction energy for the (5,5)@(17,1) DWCNT with 10\,nm and 20\,nm lengthes.}\label{pic1}
\end{figure}

\begin{figure}[!hbp]
\includegraphics[scale=0.13]{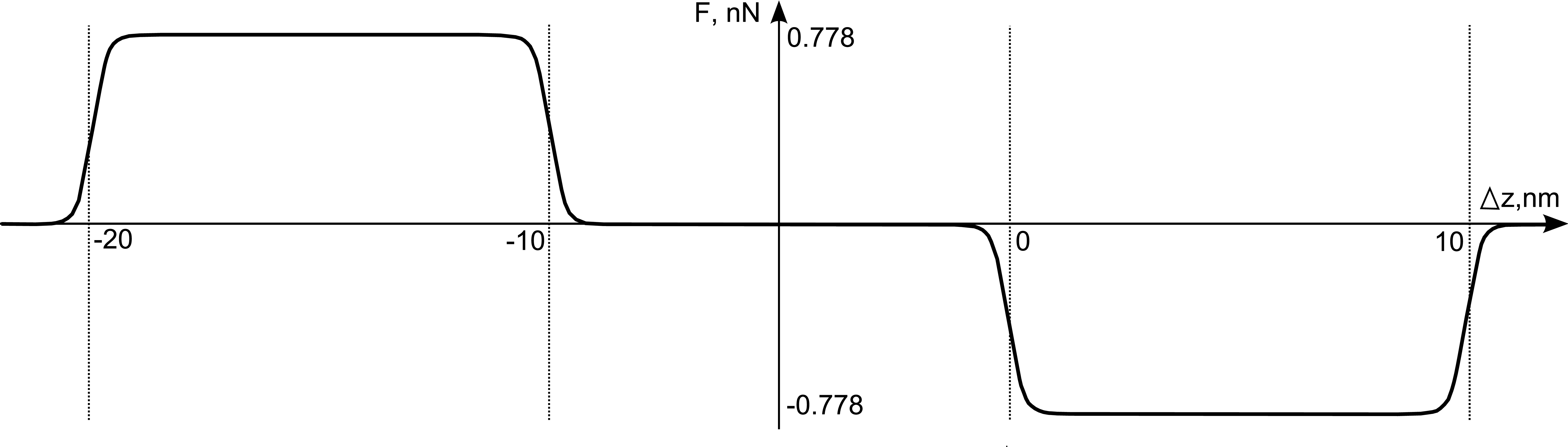}
\caption{Longitudinal intertube interaction force for (5,5)@(17,1) DWCNT with 10\,nm and 20\,nm lengthes.}\label{pic2}
\end{figure}

\section{Telescopic oscillations in DWCNT}

If the outer tube is rigidly mounted, then the longitudinal motion of the internal tube is described by Newton equation:
\begin{equation}\label{2.1}
a_{z}(\Delta z)=-\frac{1}{m}\frac{\partial U(\Delta z)}{\partial z},
\end{equation} where $a$ is the acceleration of the inner tube with mass $m$.

We ignore here the contribution of some defect-induced dissipative forces since for high-quality nanotubes they are by several orders lower than the retraction force due to self-healing mechanism \cite{zettl, zheng, kis}.

By (\ref{2.1}) the motion of inner tube is cyclic with the period
\begin{equation}\label{b.1}
\tau(E_{0})=\sqrt{2m}L \int_{L_{2}-L_{1}-\Delta z_{0}}^{\Delta z_{0}} \frac{d(\Delta z)}{\sqrt{E_{0}-U(\Delta z)}},
\end{equation}
where maximal displacement $\Delta z_{0}$ is determined by the equation
$U(\Delta z_{0}) \equiv U(L_{2}\!-\!L_{1}\!-\!\Delta z_{0})=E_{0}$.

Due to the special form of potential (\ref{1.3}) we can separate out two limiting forms of motion (Fig.\ref{pic1},\ref{pic2}):

1) steady movement for $\Delta z_{0}\gg b$ while the potential is linear in $\Delta z$;

2) small oscillations when $\Delta z_{0} \lesssim b$ and the potential is quadratic in $\Delta z$.

It is obvious that for real DWCNTs the interaction energy and force are affected by the atomic structure of its shells. As a result the interaction energy is modulated \cite{guo} with period defined by the lattice parameters of both shells. The amplitude of energy modulation can reach a value of 1000 K for zigzag@zigzag and $60\!-\!100$ K for armchair@armchair DWCNTs (for 5 nm length inner shell) and is linear in length.

On the other hand due to incommensurability of atomic lattices for most chiral nanotubes as well as armchair@chiral or zigzag@chiral pairs the modulation period can be much bigger than the whole DWCNT length. This means that impact of the shells structure substantially reduces as the smaller nanotube length increases.

Actually the interaction between two (or more) nanotubes of different length is well-described by the continuum model if the oscillation energy is much higher than 1000 K which corresponds to the great amplitude telescopic motion ($\Delta z_{0}\gg b$). The small amplitude oscillations also can be considered within the continuum model for most cases of incommensurate nanotubes for which the energy modulation amplitude varies between $10^{-2}$ and 10 K.

Furthermore, for DWCNTs with shell of equal length the effect of lattice structure on the intertube interaction energy is negligible compared to that of nanotube edges. As a result if $L_{2}-L_{1}<0.4$ nm (where 0.4 nm is the van der Waals force saturation displacement) the continuum model is valid regardless of temperature and shells structure.

\subsection{Large-amplitude oscillations in DWCNT}

\begin{figure}[!hbp]
\includegraphics[scale=0.33]{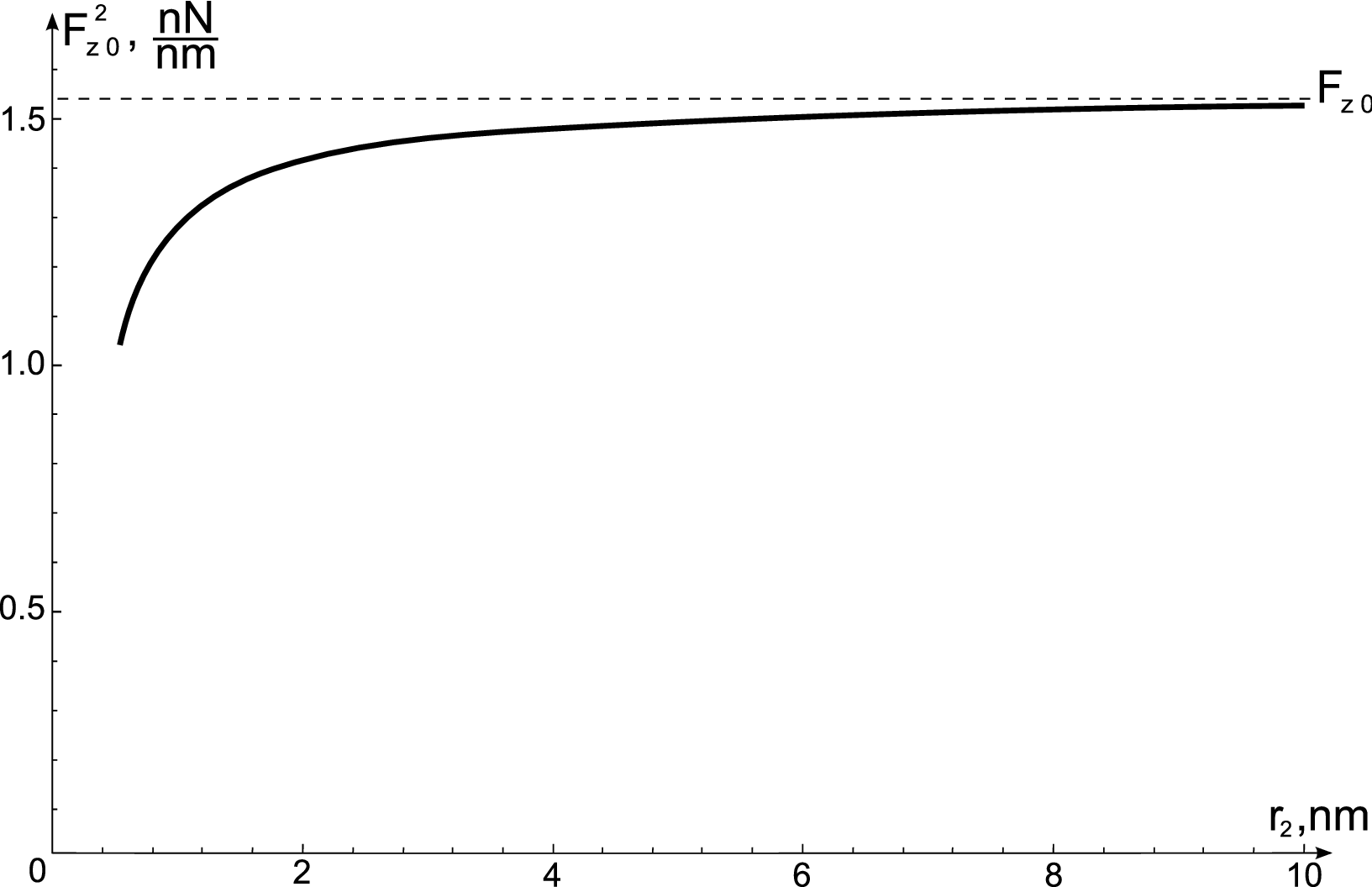}
\caption{The external shell radius dependence of $F_{z 0}^{2}$ for the natural DWCNT with interlayer distance $d=0.34$ nm.}\label{pic3}
\end{figure}

When the displacement $\Delta z_{0}$ is greater than few nanometers the potential energy is linear on $\Delta z$ except small-displacement region (with quadratic potential energy) which can be neglected. In such a case the period of oscillation can be derived from simple formulas for the steady and uniformly accelerated motion. For equal-lengths tubes the period takes the form
$$
\tau(E_{0})=4 \sqrt{\frac{2 (E_{0}-U_{min}) }{a_{z} F_{z}}}=4 \frac{\sqrt{ 2 m (E_{0}-U_{min})} }{F_{z}},
$$
where $F_{z}$ is the longitudinal component of retraction force
\begin{equation}\label{a.4}
F_{z}(r_{1},r_{2})=\frac{3\pi^{3} \sigma^{2}}{2} \, r_{1} r_{2}  \left( \frac{21}{32} \, \gamma_{12} \frac{_{2}\textrm{F}_{1} \! \left( \frac{1}{2}, \frac{11}{2}, 1,
\frac{4r_{1}r_{2}}{(r_{1}+r_{2})^{2}} \right)}{(r_{1}+r_{2})^{11}} - \gamma_{6}\frac{_{2}\textrm{F}_{1} \! \left( \frac{1}{2}, \frac{5}{2}, 1,
\frac{4r_{1}r_{2}}{(r_{1}+r_{2})^{2}} \right)}{(r_{1}+r_{2})^{5}} \right).
\end{equation}
Taking into account that \,$r_{2}\!=\!r_{1}\!+\!d$\, for natural DWCNTs, expression (\ref{a.4}) may be written as
$$
F_{z}(r_{1},r_{2})\equiv-F_{z 0}^{1}(r_{1})r_{1}\equiv-F_{z 0}^{2}(r_{2})r_{2}, \,\,\,\,\,\,\,\,
$$
where $F_{z 0}^{1}(r_{1})$ and $F_{z 0}^{2}(r_{2})$ are approximately constant for tubes of rather large radii and their asymptotic value $F_{z 0}$ is about $1.54 \frac{\textrm{nN}}{\textrm{nm}}$ (Fig.\ref{pic3}).

In terms of maximal displacement $\Delta z_0=(E_{0}-U_{min})|F_{z}|^{-1}$ the period can be rewritten as follows
$$
\tau=4 \sqrt{\frac{2\Delta z_{0}}{|a_{z}|}}=4 \sqrt{\frac{2 m \Delta z_{0}}{|F_{z}|}}.
$$

If $L_{2}>L_{1}$ then the region of steady motion also contributes to (\ref{b.1})
$$
\tau=\tau_{accelerated}+\tau_{steady}\!= 4 \frac{\sqrt{ 2 m (E_{0}-U_{min})} }{|F_{z}|} \left(1 {+} \frac{|F_{z}|}{4} \, \frac{L_{2}{-}L_{1}}{E_{0}{-}U_{min}} \right)\!
=4\sqrt{\frac{2 m \Delta z_{0}}{|F_{z}|}} \left(1\!+\!\frac{L_{2}\!-\!L_{1}\!}{4\Delta z_{0}} \right).
$$
Actually the oscillatory period does not depend on inner tube radius $r_{1}$ (if it is sufficiently large) since both the inner tube mass $m$ and retraction force $F_{z}$ are linear with $r_{1}$.

If the outer tube is also mobile, then the above expressions for periods remain to be valid with $m$ replaced by the reduced mass
$$M=\frac{m_{1}m_{2}}{m_{1}+m_{2}}.$$
Note that the interaction energy of atoms forming the tubes rapidly decreases with the interatomic distance. Therefore it is enough to consider only interaction of adjacent tubes in MWCNT. Since some adjacent shells of MWCNT can be rigidly glued by defects, then glued tubes should be considered as double-sided shells with integrated masses.

\subsection{Thermal oscillations of DWCNT}

For low temperatures the telescopic oscillations are the smallest frequency 1D modes in DWCNT. Therefore for $T\rightarrow 0$ by Boltzmann theorem
their mean energy is $\bar{E}=k_{B}T.$

For small oscillations the maximal potential energy of DWCNT coincides with $\bar{E}$:
$$ 
U_{max}=\frac{k (\Delta z_{0})^{2}}{2}=\bar{E},
$$ 
where $k$ is the rigidity parameter. Taking into account that in harmonic approximation the rigidity is the second derivative of potential energy
on the inner tube longitudinal displacement and assuming the tube radius is much smaller of its length we obtain for $L_{1}=L_{2}$ the following expression
$$ 
k(r_{1},r_{2})=4\pi\sigma^{2} r_{1} r_{2} \int_{0}^{2\pi}\left( \frac{\gamma_{6}}{ (r_{1}^{2}+r_{2}^{2}-2r_{1}r_{2}\cos{\theta})^{3}} -
\frac{\gamma_{12}}{(r_{1}^{2}+r_{2}^{2}-2r_{1}r_{2}\cos{\theta})^{6}} \right)d\theta.
$$ 

The harmonic oscillations frequency for this $k(r_{1},r_{2})$ is
$$ 
\omega_{0}=\frac{1}{2\pi}\sqrt{\frac{k}{m}},
$$ 
and the amplitude of longitudinal thermal oscillations can be estimated using the next relation
$$ 
\frac{\Delta z_{0}}{\sqrt{T}}=\sqrt{\frac{2k_{B}}{k}}\approx \sqrt{6}\cdot 10^{-3} \,\,  \frac{\textrm{nm}}{\sqrt{\textrm{K}}}.
$$ 
It can be shown that $\Delta z_{0}$ is few times smaller than graphene lattice parameter even for $T\sim300$ K.

To model the intertube interaction force $F(r_{1},r_{2},L_{1},L_{2},\Delta z)$ in the case of $L_{1}\neq L_{2}$ depending on the inner tube edge position $\Delta z$ let us assume that the axis of outer tube coincides with the interval $[0,L_{2}]$ of real axis and introduce two parameters:
$$
F_{0}(r_{1},r_{2},L_{1},L_{2})=\frac{3\pi^{3}\!\sigma^{2}}{4}\,r_{1} r_{2}\,\, \textrm{sgn}(L_{2}-L_{1})
  \left(\!\gamma_{12} \frac{21}{32}\frac{\,_{2}\textrm{F}_{1}\!\left(\!\frac{1}{2},\!\frac{11}{2},\!1,\!\frac{4r_{1}r_{2}}{(r_{1}+r_{2})^{2}} \right)}{(r_{1}+r_{2})^{11}}
  -\gamma_{6} \frac{\,_{2}\textrm{F}_{1}\!\left(\!\frac{1}{2},\!\frac{5}{2},\!1,\!\frac{4r_{1}r_{2}}{(r_{1}+r_{2})^{2}} \right)}{(r_{1}+r_{2})^{5}}\right),
$$
\begin{multline}\label{3.8}
k(r_{1},r_{2},L_{1},L_{2})=4 \pi^{2} r_{1} r_{2} \sigma^{2} \left( \gamma_{12}\!
 \left[
   \frac{\,_{2}\textrm{F}_{1}\!\left(\!\frac{1}{2},\!\frac{12}{2},\!1,\!\frac{4r_{1}r_{2}}{(r_{1}+r_{2})^{2}} \right)}{(r_{1}+r_{2})^{12}}+
   \frac{\,_{2}\textrm{F}_{1}\!\left(\!\frac{1}{2},\!\frac{12}{2},\!1,\!
   \frac{4r_{1}r_{2}}{(r_{1}+r_{2})^{2} + (L_{1}-L_{2})^{2}} \right)}{[(r_{1}+r_{2})^{2} + (L_{1}-L_{2})^{2}]^6}
 \right] \right. \\
 \left. -\gamma_{6}
 \left[
   \frac{\,_{2}\textrm{F}_{1}\!\left( \frac{1}{2}, \frac{6}{2}, 1, \frac{4r_{1}r_{2}}{(r_{1}+r_{2})^{2}} \right)}{(r_{1}+r_{2})^{6}}+
   \frac{\,_{2}\textrm{F}_{1}\!\left( \frac{1}{2}, \frac{6}{2}, 1, \frac{4r_{1}r_{2}}{(r_{1}+r_{2})^{2}+(L_{1}-L_{2})^{2}} \right)}{[(r_{1}+r_{2})^{2}+(L_{1}-L_{2})^{2}]^3}
 \right]
\right)=k_{0}r_{2},
\end{multline}
where $k_{0}$ is almost independent of $r_{1},r_{2}$ for rather large values of these parameters. For $r_{1}>10$ nm we have $k_{0}\approx -3.7 \, \frac{\textrm{nN}}{\, \textrm{nm}^2}$.

For small maximum retractions of inner tube ($\Delta z\!<0.3$ nm) $F(r_{1},r_{2},L_{1},L_{2},\Delta z)$ can be written as follows
$$ 
F(r_{1},r_{2},L_{1},L_{2},\Delta z) =
\begin{cases}
F_{0} + k \Delta z, & \text{$0.3\,\textrm{nm}<\Delta z<x_{0}$} \\
0, &\text{$x_{0}<\Delta z<L_{2}-L_{1}-x_{0}$} \\
-F_{0}\! + k\,{\cdot}(L_{2}{-}L_{1}{-}\Delta z), & \text{$L_{2}{-}L_{1}{-}x_{0}\!<\!\Delta z\!<\!L_{2}{-}L_{1}\!+0.3\,\textrm{nm}$}\\
\end{cases}
$$ 

Here $x_{0}=\left|\frac{F_0}{k} \right|$ is the displacement, which makes the longitudinal retraction force equals to zero,
$k(r_{1},r_{2},L_{1},L_{2})$ is the DWCNT longitudinal rigidity and $F_{0}(r_{1},r_{2},L_{1},L_{2})=F(r_{1},r_{2},L_{1},L_{2},0)$.

In case of tubes with significantly different lengths ($|L_1-L_2| \gtrsim 1$ nm) the expression (\ref{3.8}) takes the form
$$ 
  k_{L_1\neq L_2}(r_{1},r_{2})=4 \pi^{2} r_{1} r_{2} \sigma^{2}
  \left( \gamma_{12} \frac{\,_{2}\textrm{F}_{1} \left( \frac{1}{2}, \frac{12}{2}, 1, \frac{4r_{1}r_{2}}{(r_{1}+r_{2})^{2}} \right)}{(r_{1}+r_{2})^{12}}
  -\gamma_{6} \frac{\,_{2}\textrm{F}_{1} \left( \frac{1}{2}, \frac{6}{2}, 1, \frac{4r_{1}r_{2}}{(r_{1}+r_{2})^{2}} \right)}{(r_{1}+r_{2})^{6}} \right)
$$ 
and in the case of equal lengths $k_{L_1=L_2}(r_{1},r_{2})=2 k_{L_1\neq L_2}(r_{1},r_{2})$.

The oscillation cycle can be considered as
$$ 
\tau=\tau_{steady}+\tau_{accelerated}=2\frac{|L_2-L_1|-2 x_0}{V_{max}}+2\pi\sqrt\frac{m}{k},
$$ 
where $V_{max}=\sqrt{\frac{2}{m}(E(\Delta z_0)-U_{min})}$ is the maximum inner tube velocity and $U_{min}=U\left(\frac{-|L_2-L_1|}{2}\right)$ is the system minimum potential energy defined by (\ref{1.5}).
With an accuracy of several percent previous equation can by approximated by
$$ 
\tau=2\sqrt{\frac{m}{k}}\left( \frac{|L_2-L_1| - 2x_0}{|\Delta z_0 + x_0|} +\pi \right).
$$ 

Since the longitudinal rigidity depends only on the tubes radii and the inner tube mass is proportional to the product of its length and radius, then the harmonic oscillation frequency scales on the length as \, $\omega_{0}\sim L^{-1/2}.$

From the above discussion it is clear that for $T\!\lesssim\!300$ K the continuum model is valid for the DWCNTs with incommensurate shells but ceases to be true for commensurate (armchair@armchair, zigzag@zigzag) and some quasi-commensurate configurations. As an example, for the majority of zigzag@chiral and armchair@chiral DWCNTs the considered thermal oscillations are possible for temperatures higher than $0.01\!-1\!$ K. As for the chiral@chiral configurations, in some cases the shells atomic structure impact may becomes negligible even for $T\sim10^{-3}$ K.

\subsection{Thermal oscillation frequencies in multi-walled CNTs}

Considering the long-amplitude oscillations of multi-walled nanotube we assumed that some nanotube's shells can be bounded by the defects, but in the case of thermal oscillations we should take into account the motion of all shells because amplitudes of their oscillations are of the same order and much lesser then interatomic distance. For simplicity assume that all MWCNT's shells are equal in length.

We obtained the MWCNT's thermal oscillations frequencies by solving the system of equations for longitudinal displacements of tubes with forces defined by the expression above for the intertube potential. In the continuum model arbitrary multi-walled nanotube can be characterized by the
inner shell radius $r_0$, the number of shells $n$ under consideration and the constant distance between adjacent shells ($d=0.34$ nm).

\begin{figure}[!hbp]
\includegraphics[scale=0.33]{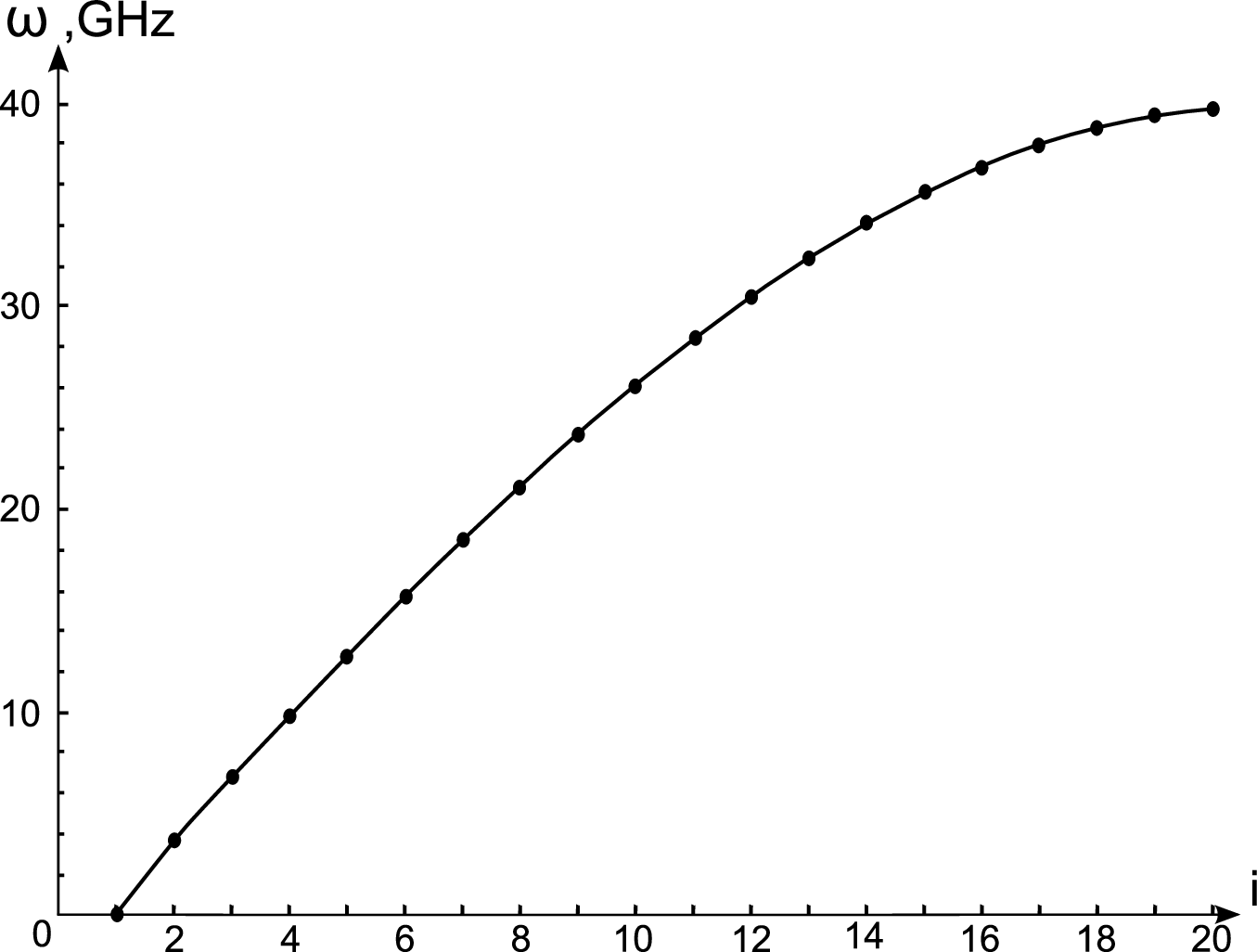}
\caption{Frequencies of small telescopic oscillations for the 50\,nm-length MWCNT with 20 shells.}\label{pic4}
\end{figure}

\begin{figure}[!hbp]
\includegraphics[scale=0.33]{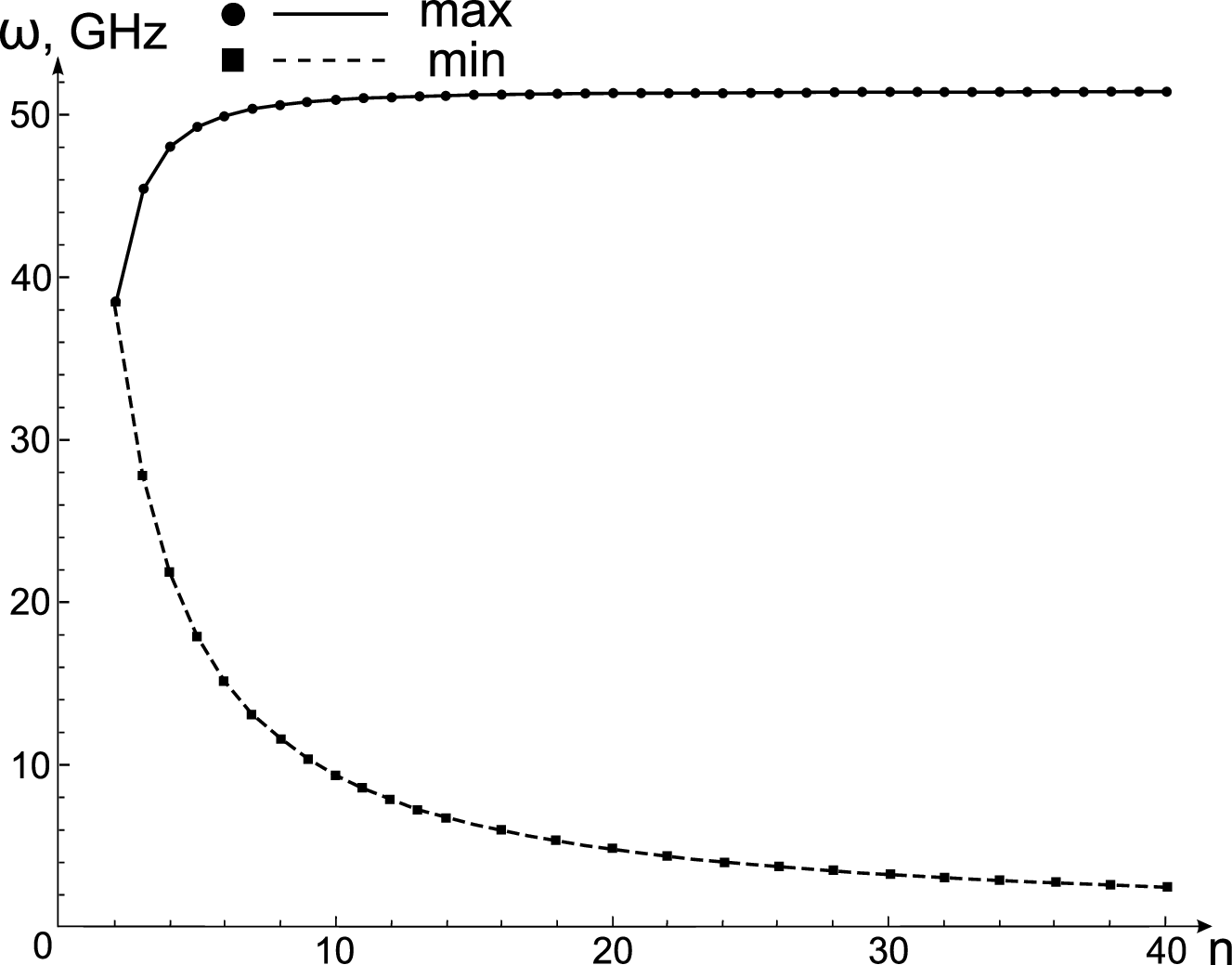}
\caption{The maximal and minimal oscillation frequency for 30\,nm-length MWCNT for different number of shells.}\label{pic5}
\end{figure}

Considering only adjacent shells interaction we find explicit values of the consequent MWCNT eigenfrequencies $\omega_i, i=\overline{1..n}$ (fig. \ref{pic4}). The smallest eigenvalue is always equal to zero corresponding to the whole nanotube translational motion. The analysis shows that the maximal frequency depends on the MWCNT characteristics but in case of tube with large number of shells ($n \gtrsim 10$) it tends to the asymptotic value which depends only on the tubes length (fig. \ref{pic5}):
$$\omega_{max} = \frac{280}{\sqrt{L}} \,\textrm{GHz}.$$

The obtained value of maximal frequency is underestimated for real tubes because defects may increase the longitudinal rigidity of MWCNT. The minimal
oscillation frequency strongly depends on the number of shells (as a result of increasing of the outer shell mass) and can be found using the following interpolation formula (fig. \ref{pic5})
\begin{equation}\label{4.4}
 \omega_{min} = \frac{4.46\times10^{11}}{\sqrt{L}} \frac{e^{-0.0128 n}}{n^{0.9365}} \,\textrm{GHz}.
\end{equation}

Using obtained frequencies the contribution of tube's telescopic oscillations to the individual MWCNT internal energy is calculated
$$ 
E(T)=\sum_{\omega_i\neq 0}\hbar\,\omega_i \left(\frac{1}{2} +  \frac{1}{\exp[\frac{\hbar\,\omega_i}{k_{B}T}]-1} \right),
$$ 
where $k_B$ is the Boltzmann constant and $T$ is an absolute temperature, and specific heat
\begin{equation}\label{4.5}
 C_v(T)=\frac{\partial E(T)}{\partial T}=\sum_{\omega_i\neq 0} k_B \left(\frac{ \frac{1}{2}\frac{\hbar\,\omega_i}{k_B T} } {\sinh[\frac{1}{2}\frac{\hbar\,\omega_i}{k_{B}T}]} \right)^2.
\end{equation}
By (\ref{4.5}) if $k_{B}T\ll \hbar \omega_{min}$, then
$$ 
C_v(T)\simeq k_{B} \left(\frac{\hbar\,\omega_{min}}{k_B T}\right)^2 \exp\left(-\frac{\hbar\,\omega_{min}}{k_B T}\right)
$$ 
while
$C_v(T)\simeq nk_{B}$ if $k_{B}T\gg \hbar \omega_{max}$.

It is well known that under the Debye temperature the bulk solid specific heat decreases as a cubic function of the temperature and
for the one dimensional structures such decreasing is given by the linear function.
In the case of MWCNTs the telescopic oscillation induced specific heat decreases exponentially in the range of small temperatures and the corresponding Debye-like temperature is
$$ 
T_{D}=\frac{\hbar \omega_{min}}{2 k_{B}}=\frac{1.7}{\sqrt{L}}\frac{e^{-0.0128 n}}{n^{0.9365}}\,\textrm{K},
$$ 
where $\omega_{min}$ is given by (\ref{4.4}). As a result of exponential decreasing MWCNT's specific heat may be several orders higher than that of environment for $T \gg T_D$ while for $T \ll T_D$ these values change over. For the natural carbon nanotubes $T_D$  varies in the interval ($10^{-3} - 1)$ K.

If MWCNT's shells have different lengths the hampering of low energy telescopic motion by the lattice structure causes the abrupt increase of the oscillation frequency and also leads to the specific heat exponential decreasing. Moreover for most MWCNTs the thermal oscillations freezing-out as a result of both processes takes place within the same temperature range from $10^{-3}$ to $1\!-\!10$ K.

For the temperature $T\!=\!1$ K (assuming that $T>T_D$) the telescopic oscillations contribution to the total nanotube's specific heat may run as high as 50\% for the relatively small double-walled nanotubes ($L=20-30$ nm, $r_2~1$ nm). For the ten-walled MWCNTs of length $L=50$ nm and external radius $r=2.5$ nm the telescopic oscillation specific heat is about $0.025 \frac{\textrm{J}}{\textrm{kg K}}$ and phonon contribution is in the range $0.2$ to $0.3 \,\frac{\textrm{J}}{\textrm{kg K}}$ \cite{lasjaunias}. The maximal electronic contribution for metallic SWCNT is estimated to be ten times smaller than that of lattice oscillations \cite{lasjaunias}. As for all semiconducting nanotubes (which are the majority of natural MWCNTs) the electronic specific heat is negligible.
Taking into account the 1D structure phonon specific heat linear decreasing in the considered temperature region it is reasonable to expect that for $T\lesssim10^{-1}$K (if it is higher than $T_D$) the telescopic oscillation contribution may become dominant.

\section{Summary}
The explicit expressions for longitudinal rigidities and frequencies of small and large-amplitude telescopic oscillations of DWCNT and MWCNT
were deduced in the framework of continuum  Lennard-Jones model borrowed from \cite{girifalco}. Besides the obtained frequencies of telescopic oscillations of MWCNT are in good agreement with available experimental data \cite{zettl} and results of numerical simulations \cite{rivera,galvao1,galvao2}.

For example the thermal oscillation frequency of 12.21 nm (7,0)@(9,9) DWCNT obtained in \cite{rivera} is $75\pm 8$ GHz while the considered model gives 62 GHz. The retraction force $F_z$ for the (9,0)@(12,0) DWCNT obtained in \cite{galvao1, galvao2} is 1.6 nN and using Lennard-Jones parameters from \cite{galvao1} it yields  $F_z=1.54$ nN. For (5,5)@(10,10) and (10,10)@(15,15) DWCNTs the maximum retraction forces ratio is 1.67 by our model and 1.7 in \cite{KangLee}. So the difference between frequencies calculated by our analytical formulas and those found by numerical methods with account of discrete structure of nanotubes lies within the $5\%$-range.

It is worth to be mentioned that investigation of multiwalled nanotube oscillations by using exact expression for the two-shell retraction force and longitudinal rigidity is not computationally intensive in contrast to the molecular dynamics simulations. This permits us easily to calculate (within the bounds of continuum model) all oscillation frequencies and corresponding parameters (such as specific heat) of any MWCNT regardless of number of shells and their configuration.

Therefore the considered Lennard-Jones continuum model is seemingly well suited for description of telescopic trembling of MWCNT.

\section*{Acknowledgments}
Authors are grateful to Prof. V.M.Adamyan for discussions and valuable remarks.


\section*{References}
\begin{thebibliography}{18}
\bibitem[1]{iijima} S. Iijima, {\it Helical microtubules of grafitic carbon}, Nature {\bf 354}, 56 (1991).
\bibitem[2]{girifalco} L. A. Girifalco, Miroslav Hodak, and Roland S. Lee, Phys. Rev. B {\bf 62}, 013104 (2000).
\bibitem[3]{zettl} J. Cumings, A. Zettl, {\it Low-Friction Nanoscale Linear Bearing Realized from Multiwall Carbon Nanotubes},
Science {\bf 289}, 602 (2000).
\bibitem[4]{treasy} M.M.J. Treasy, T.W. Ebbesen, J.M. Gibson, {\it Exceptionally high Young's modulus observed for individual carbon nanotubes},
Nature {\bf 381}, 678 (1996).
\bibitem[5]{zheng} Q. Zheng, Q. Jiang, Phys. Rev. Lett. {\bf 88}, 045503 (2002).
\bibitem[6]{forro} L. Forro, {\it Nanotechnology: Beyond Gedanken Experiments}, Science {\bf 289}, 560 (2000).
\bibitem[7]{strastava} D.W. Srivastava, {\it A phenomenological model of the rotation dynamics of carbon nanotube gears with laser electric fields},
Nanotechnology {\bf 8}, 186 (1997).
\bibitem[8]{tu} Z.C. Tu, X. Hu, {\it Molecular motor constructed from a double-walled carbon nanotube driven by axially varying voltage},
Phys. Rev. B {\bf 72}, 033404 (2005).
\bibitem[9]{baowan1} Duangkamon Baowan and James M. Hill, {\it Force distribution for double-walled carbon nanotubes and gigahertz oscillators},
Z. angew. Math. Phys. {\bf 58}, 857–875 (2007).
\bibitem[10]{baowan2} Duangkamon Baowan, Ngamta Thamwattana, James M. Hill, {\it Suction energy and offset configuration for double-walled carbon nanotubes}, Communications in Nonlinear Science and Numerical Simulation {\bf 13}, 1431–1447 (2008).
\bibitem[11]{kiang} C.H. Kiang, M. Endo, P.M. Ajayan, G. Dresselhaus, M.S. Dresselhaus, {\it Size effects in carbon nanotubes}, Phys. Rev. Lett. {\bf 81}, 1869 (1998).
\bibitem[12]{kis} A. Kis, K. Jensen, S. Aloni, W. Mickelson, and A. Zettl, {\it Interlayer forces and ultralow sliding friction in multiwalled carbon nanotubes.} Phys. Rev. Lett. {\bf 97}, 025501 (2006).
\bibitem[13]{guo} W. Guo, Y. Guo, H. Gao, Q. Zheng, and W. Zhong, {\it Energy Dissipation in Gigahertz Oscillators from Multiwalled Carbon Nanotubes},
Phys. Rev. Lett. {\bf 91}, 125501 (2003).
\bibitem[14]{lasjaunias} J.C. Lasjaunias, K. Biljakovi\'{c}, Z. Benes, J.E. Fischer, and P. Monceau, {\it Low-temperature specific heat of single-wall carbon nanotubes}, Phys. Rev. B {\bf 65}, 113409 (2002)
\bibitem[15]{rivera} Jose L Rivera, Clare McCabe and Peter T Cummings, {\it The oscillatory damped behaviour of incommensurate double-walled carbon nanotubes}, Nanotechnology {\bf 16},  186–198 (2005).
\bibitem[16]{galvao1} S.B. Legoas, V.R. Coluci, S.F. Braga, P.Z. Coura, S.O. Dantas and D.S. Galvao, {\it Gigahertz nanomechanical oscillators based on carbon nanotubes}, Nanotechnology {\bf 15},  S184–S189 (2004).
\bibitem[17]{galvao2} S.B. Legoas, V.R. Coluci, S.F. Braga, P.Z. Coura, S.O. Dantas, and D.S. Galvao, {\it Molecular-Dynamics Simulations of Carbon Nanotubes as Gigahertz Oscillators}, Phys. Rev. Lett. {\bf 90}, 055504 (2003).
\bibitem[18]{KangLee} J.W. Kang and J.H. Lee, {\it Frequency characteristics of triple-walled carbon nanotube gigahertz devices},
Nanotechnology {\bf 19},  285704 
 (2008).

\end{thebibliography}
\end{document}